%% file: main.tex
\documentclass[a4paper,11pt,dvipsnames]{article}
\usepackage{pos,tikz,subcaption}
\usetikzlibrary{arrows.meta, decorations.pathmorphing, decorations.markings,bending,positioning}

\newcommand{\rcirclearrow}{\circlearrowleft}

\title{New Tools in the Landau Bootstrap}

\author{Piotr Bargie\l{}a}
\author{Giulio Crisanti}
\author{Craig Larkin}
\author{Luke Lippstreu}
\author*{Andrew J.~McLeod}
\author{Maria Polackova}
\author{Laura Walsh}

\affiliation{Higgs Centre for Theoretical Physics, 
    School of Physics and Astronomy, 
    The University of Edinburgh 
    \\ Edinburgh EH9 3FD, Scotland, UK}

\emailAdd{pbargiel@ed.ac.uk}
\emailAdd{g.crisanti@ed.ac.uk}
\emailAdd{C.D.Larkin@sms.ed.ac.uk}
\emailAdd{llippstr@ed.ac.uk}
\emailAdd{andrew.mcleod@ed.ac.uk}
\emailAdd{maria.polackova@ed.ac.uk}
\emailAdd{L.E.Walsh-1@sms.ed.ac.uk}

\abstract{We describe recent advances in our understanding of the analytic structure of Feynman integrals. In particular, we describe two new classes of constraints on such integrals, that identify discontinuities that either cannot be repeated, or that always give rise to the same result (no matter which other discontinuities are computed first). These new constraints hold at all orders in dimensional regularization, and provide us with new input for the Landau bootstrap, where information about the singularities and discontinuities of individual Feynman integrals is used to construct their functional form.}

\FullConference{
Loops and Legs in Quantum Field Theory (LL2026) \\  
12-17, April, 2026 \\
Bayreuth, Germany
}

\begin{document}
\maketitle

\section{Introduction}

Methods for studying the analytic structure of perturbative quantum field theory have developed rapidly in recent years, driven in part by the novel application of tools from twisted intersection theory and Picard--Lefschetz theory~\cite{Berghoff:2022mqu,Britto:2023rig,Hannesdottir:2024hke,Muhlbauer:2022ylo,Frellesvig:2019uqt,Mizera:2017rqa,Mandal:2024wun,Mizera:2019gea}. In particular, the locations at which singularities and discontinuities appear in perturbative scattering amplitudes (and related quantities in quantum field theory) have long been known to obey constraints dictated by locality, unitarity, and crossing symmetry. These constraints are reflected in the geometry of the differential forms and integration contours that enter these quantities, and can thus be probed using tools from algebraic geometry. 

Understanding these physical constraints is especially useful in perturbative bootstrap methods, where scattering amplitudes (or similar quantities) are determined by building all of the mathematical properties a given example is expected to exhibit into an ansatz~\cite{Caron-Huot:2020bkp,Arkani-Hamed:2022rwr}. When such bootstrap methods can be leveraged, they allow us to sidestep more computationally prohibitive approaches, such as evaluation via direct integration or differential equation methods. In the case of individual Feynman integrals, this bootstrap strategy can be made especially concrete. The loci where singularities can arise in these integrals are governed by the Landau equations, which identify where the integration contour can be pinched by the singularities of the integrand~\cite{nakanishi1959,Landau:1959fi,Bjorken:1959fd}. The \emph{Landau Bootstrap}~\cite{Hannesdottir:2024hke} then pairs this information with knowledge of the types of special functions that are expected to arise, and which sequences of discontinuities are consistent with physical principles. Jointly, these constraints identify a finite space of functions that one's integral is expected to evaluate to. The functional form of the integral can then be determined using information about its behavior in special kinematic limits. 

In these proceedings, we focus on recent advances within the Landau bootstrap program for predicting what sequences of discontinuities can appear in Feynman integrals. In the case of integrals that evaluate to multiple polylogarithms~\cite{Chen,G91b,Goncharov:1998kja,Remiddi:1999ew,Borwein:1999js,Moch:2001zr}, this question can be phrased directly in terms of what sequences of symbol letters (or logarithmic branch points) are allowed to appear in an integral's symbol~\cite{Goncharov:2010jf}. In Feynman integrals that give rise to functions that go beyond multiple polylogarithms~\cite{Bourjaily:2022bwx}, the implications of these analytic constraints become more opaque; however, the same question remains pertinent: after computing a given sequence of discontinuities, what singularities and discontinuities can still arise?

One way to answer this question is to track what happens to the integration contour itself. To see this, consider the discontinuity of an integral $\mathcal{I}$ with respect to a branch cut that starts at $\lambda=0$. This discontinuity can be computed as the difference between the value of $\mathcal{I}$ before and after it has been analytically continued around the branch point at $\lambda=0$:
\begin{equation} \label{eq:disc_def}
    \text{Disc}_\lambda \mathcal{I} = \left(\rcirclearrow_\lambda - 1 \right) \mathcal{I}\, .
\end{equation}
In general, this difference can be written as an integral over the original integrand but involving a new integration contour. Picard--Lefschetz theory gives us a precise formulation of this statement. 
Namely, the discontinuity in~\eqref{eq:disc_def} can be computed from specific pieces of (local) homological data: a vanishing cycle, an associated vanishing cell, and the intersection number of this vanishing cell with the original contour~\cite{Berghoff:2022mqu,pham2011singularities}. 
Although working out this homological information in full detail is prohibitively difficult in most examples, nontrivial constraints can still be placed on the discontinuities of $\mathcal{I}$ just using information about what integration endpoints appear in the vanishing cycle (considered in Feynman parameter space)~\cite{Berghoff:2022mqu,Britto:2023rig,Hannesdottir:2024cnn}. In practice, this amounts to computing the Euler characteristic of different topological spaces that can be associated with $\mathcal{I}$ and its discontinuities. More specifically, the Euler characteristic should drop in kinematic configurations where the integral can become singular~\cite{Fevola:2023kaw}---thus, by checking whether the Euler characteristic still drops after we have computed a sequence of discontinuities, we can place \emph{genealogical constraints} on the ordered sequences of discontinuities that can appear~\cite{Hannesdottir:2024cnn}.

In what follows, we highlight two new applications of these Picard--Lefschetz-inspired ideas. First, we describe how certain \emph{non-repeating discontinuities} effectively erase their own possibility. Namely, after a discontinuity has been computed with respect to one of these singularities, it is no longer possible for a singularity to arise in the same kinematic location (on any Riemann sheet).\footnote{In language that harkens back to the nineteenth century, we might say that these singularities can only be realized by negating their immediate form~\cite{hegel1979phenomenology}.} Second, we propose a \emph{Lefschetz uniqueness principle}, which identifies certain discontinuities that must always return the same result (considered to all orders in dimensional regularization), up to a numerical proportionality coefficient, whenever they are computed. We finish by highlighting progress that has also been made towards understanding when the Steinmann relations are violated in massless Feynman integrals~\cite{Hannesdottir:2025bss}.


\section{The Picard--Lefschetz Formula and Hierarchical Constraints}
The key statement that our results build on is the Picard--Lefschetz formula. This formula expresses the discontinuity of an integral with respect to a branch cut starting at $\lambda=0$ as a sum of integrals involving the same differential form, but new integration contours $\nu_i$ referred to as vanishing cycles. Each of these new integrals is also multiplied by an intersection number, giving
\begin{equation}\label{PL_formula}
    \text{Disc}_{\lambda=0}\left( \int_{\sigma} \omega \right) = \sum_{i} \langle \tilde{\nu}_{i} | \sigma \rangle \int_{\nu_{i}} \omega\, ,
\end{equation}
where $\langle \tilde{\nu}_{i} | \sigma \rangle$ represents the oriented intersection number between the original integration contour $\sigma$ and the vanishing cell $\tilde{\nu}_i$ that is dual to the vanishing cycle $\nu_i$. (For more details, see~\cite{pham:1965}.) Importantly, $\tilde{\nu}_i$ and $\nu_i$ have the property that they become trivial in their respective homology groups in the strict $\lambda \to 0$ limit. Thus, for a nontrivial discontinuity to exist about $\lambda=0$, it must be that the dimension of the homology group that represents the set of possible integration contours drops when $\lambda=0$. Otherwise, there are no candidate vanishing cycles that can appear in equation~\eqref{PL_formula}. 

In practice, this drop in the dimension of the relevant homology group can be detected by computing the signed Euler characteristic $\chi$ of an appropriate topological space, constructed out of the (co)homological data of $\mathcal{I}$~\cite{pham,Fevola:2023kaw}. For instance, in the Lee-Pomeransky representation 
\begin{equation} \label{eq:lee_pom}
    \mathcal{I} \sim \int_{\alpha_i \ge 0} \frac{\prod_i^n d \alpha_i}{\mathcal{G}^{\frac{D}{2}}} \, ,
\end{equation}
where $\mathcal{G}$ is a polynomial of the external kinematic variables and the Feynman parameters $\alpha_i$, the relevant topological space is
\begin{align} 
    X = \mathbb{C}^n\, \setminus \left\{\mathcal{G} = 0 \, \bigcup_i^n \alpha_i = 0 \right\} \, ,
\end{align}
namely the space of Feynman parameters in which the singular locus $\mathcal{G} = 0$ and the endpoints $\alpha_i=0$ have been excised. The Euler characteristic of this space, $\chi(X)$, counts the number of independent contours the differential form in~\eqref{eq:lee_pom} can be integrated over, which (as per the above reasoning) should drop if it is possible for a discontinuity to arise in the $\lambda \to 0$ limit. In practice, $\chi(X)$ can be computed by finding the number of zero-dimensional solutions to the so-called critical point equations:
\begin{equation}\label{eq:critical-point-equations}
 d \log ({\alpha_1}^{\nu_1}\cdots {\alpha_n}^{\nu_n} \mathcal{G}^{-D/2}) =0\,,
\end{equation}
where each $\nu_i$ is a generic complex parameter. For more details, see~\cite{Frellesvig:2019kgj,Fevola:2023fzn,lee_critical_2013,Frellesvig:2019uqt,Huh_2013,matsubaraheo2025hypergeometricdiscriminants}.

To get constraints on the sequential discontinuities of $\mathcal{I}$, we need just one further observation---that the relevant topological space $X$ can change after each discontinuity is computed, since each time the new integration contour(s) in the Picard--Lefschetz formula have the potential to involve fewer $\alpha_i=0$ endpoints. Which endpoints disappear can be determined by checking which sectors lose critical points in the corresponding singular limit (more on this below). By probing for drops of the Euler characteristic in the modified spaces
\begin{align} 
    X_E = \mathbb{C}^n\, \setminus \left\{\mathcal{G} = 0 \, \bigcup_{i\in E} \alpha_i = 0 \right\} \label{eq:X_E}
\end{align}
that retain information just about the endpoints at $\alpha_i =0$ for $i \in E$, we can derive \emph{genealogical constraints}. These constraints take the form
\begin{align}  \label{eq:genealogical} \text{Disc}_{\lambda_1 = 0} \cdots \text{Disc}_{\lambda_2 = 0} \cdots \mathcal{I} = 0 \, ,
\end{align}
where the dots indicate that any sequence of further discontinuities is allowed to appear. For more details, see~\cite{Hannesdottir:2024cnn,to_appear}.

\section{Non-Repeating Discontinuities}
\label{sec:non_repeating_letters}

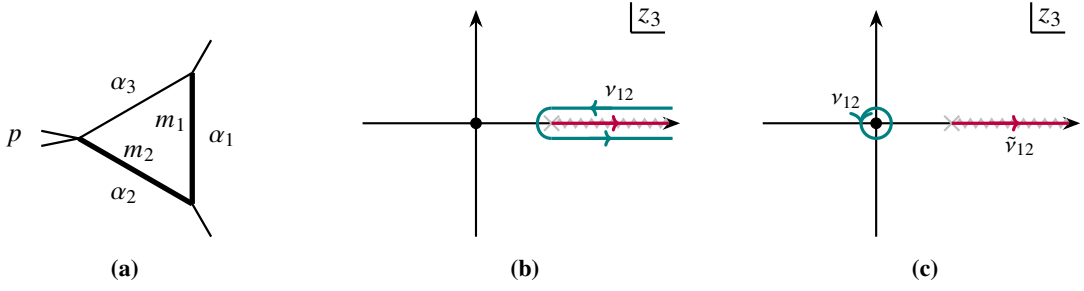
\begin{figure}[t]
    \centering
    \begin{subfigure}[b]{0.3\textwidth}
        \centering
        \input{tikz_figs/1a.tex}
        \caption{}
        \label{subfig:left}
    \end{subfigure}
    \hfill 
    \begin{subfigure}[b]{0.3\textwidth}
        \centering
        \input{tikz_figs/1b.tex}
        \caption{}
        \label{subfig:middle}
    \end{subfigure}
    \hfill 
    \begin{subfigure}[b]{0.3\textwidth}
        \centering
        \input{tikz_figs/1c.tex}
        \caption{}
        \label{subfig:right}
    \end{subfigure}

    \caption{\textbf{(a)} Hooked triangle Feynman diagram with $p^2 \neq 0$ and internal masses $m_1^2$ and $m_2^2$. \textbf{(b)} The $z_3$ integration plane after we have computed discontinuities about $m_1^2=0$ and then $(m_1^2-m_2^2)=0$. \textbf{(c)} By contour deformation, $\nu_{12}$ is equivalent to a residue contour around the simple pole at the origin. }
    \label{fig:hooked_triangle}
\end{figure}

While the methods introduced in~\cite{Hannesdottir:2024cnn} can be used to derive constraints on pairs of discontinuities that begin at different kinematic loci $\lambda_1=0$ and $\lambda_2=0$, restrictions can also be placed on repeated discontinuities with respect to a single kinematic locus $\lambda=0$:
\begin{align}
\text{Disc}_{\lambda = 0} \cdots \text{Disc}_{\lambda = 0} \cdots \mathcal{I} = 0 \, , \label{eq:non_repeating_constraint}
\end{align}
To illustrate this, we consider the hooked triangle integral $\mathcal{I}_{\text{hook}}$ depicted in Figure~\ref{subfig:left} (recently also considered in~\cite{Britto:2025wzt}). This integral has singularities at six locations in the space of external kinematics, namely at the vanishing loci of the polynomials
\begin{equation} \label{sings}
   \{m_1^2 \,,\, m_2^2 \,,\, p^2 \,,\, p^2-m_1^2 \,,\, m_1^2-m_2^2-p^2 \,,\, m_1^2-m_2^2\} \,.
\end{equation}
We will primarily be interested in the last of these singularities, which arises when $m_1 \to m_2$. This singularity does not appear on the physical Riemann sheet, but is present in the discontinuities of $\mathcal{I}_{\text{hook}}$ about either $m_1^2=0$ or $m_2^2=0$. 

First, let us analyze the solutions to the critical point equations~\eqref{eq:critical-point-equations}. Making use of the $\mathcal{G}$ polynomial for the hooked triangle,
\begin{equation}
    \mathcal{G}_{\text{hook}}=\alpha_1 + \alpha_2 + \alpha_3 - p^2 \alpha_1 \alpha_3 +(m_1^2 \alpha_1 +m_2^2 \alpha_2)(\alpha_1 + \alpha_2 + \alpha_3) \, ,
\end{equation}
we find four solutions for generic values of $p^2$, $m_1^2$, and $m_2^2$. These solutions all appear in different sectors, namely
\begin{equation}
    \{\alpha_1\},\quad
    \{\alpha_2\},\quad
    \{\alpha_1,\alpha_3\},\quad
\{\alpha_1,\alpha_2,\alpha_3\}\, ,
\label{sectors}
\end{equation}
where each sector is labeled by the Feynman parameters that are not set to their endpoint value (so, for instance, sector $\{\alpha_1\}$ corresponds to $E = \{2,3\}$ in~\eqref{eq:X_E}; in momentum space, it is the sector in which just the propagator associated with $\alpha_1$ is on shell). 

In general, solutions to the critical point equations can be used to construct a basis of contours over which the integral~\eqref{eq:lee_pom} can be integrated (the basis of Lefschetz thimbles). As explained in the last section, this means we expect at least one of these critical point solutions (and the corresponding integration contour) to vanish on the support of each of the singular loci in~\eqref{sings}, in order for there to be a candidate vanishing cell in the Picard--Lefschetz formula. Moreover, this reasoning can be applied sector by sector. In this case, this means that the critical point(s) that must vanish for each singularity can be easily deduced, since there exists a unique critical point in each sector. For instance, only a single solution to the Landau equation sets $m_1^2=0$; as it involves putting the propagator with mass $m_1$ on shell while keeping the other two propagators off shell, the critical point that vanishes must be in the $\{\alpha_1\}$ sector. 

This expectation is indeed borne out for the first five singularities in~\eqref{sings}. However, when one sets $(m_1^2-m_2^2) = 0$, something more interesting happens. Instead of any of the existing critical point solutions vanishing, a \emph{new} solution to the critical point equations arises in sector $\{ \alpha_1, \alpha_2 \}$. This is, in some sense, not unexpected, since this matches the sector where the $(m_1^2-m_2^2) = 0$ solution to the Landau equations is found (which requires doing an appropriate blowup~\cite{Fevola:2023kaw}). However, this solution is strange insofar as it is one-dimensional (as opposed to a point). The homological interpretation of this phenomenon is subtle, and we defer discussion of it to~\cite{to_appear}. For now, we merely note that \emph{something} interesting happens to the topology of $X_{3}$ in the $m_1 \to m_2$ limit, and thus it is not unusual for a singularity to have developed here. However, we still need to identify a vanishing cycle that can appear in the Picard--Lefschetz formula that is well-defined in the neighborhood of this singularity (and not just in the strict limit). Since only a single critical point exists in generic kinematics once the endpoints at $\alpha_1=0$ and $\alpha_2=0$ have been forgotten (or the propagators associated with $\alpha_1$ and $\alpha_2$ have been put on shell), there is only a single candidate. The vanishing cycle must be the one that appears in sector $\{\alpha_1,\alpha_2,\alpha_3\}$.

Now comes the key point. The Lefschetz thimble that one can construct starting from the critical point in the $\{\alpha_1,\alpha_2,\alpha_3\}$ sector no longer ends on any of the $\alpha_i=0$ boundaries found in the original Feynman integral. In momentum space, this translates to the statement that all three propagators are on shell. At the same time, reaching the $(m_1^2-m_2^2) = 0$ solution to the Landau equations requires setting $\alpha_3$ to its endpoint value (meaning the propagator associated with $\alpha_3$ must be off shell). It follows that there can no longer be any singularity in the $m_1 \to m_2$ limit once we have applied the operator $\text{Disc}_{(m_1^2-m_2^2) = 0}$. This translates to a constraint of the form~\eqref{eq:non_repeating_constraint}. 

We can see how this phenomenon comes about more concretely in the Baikov representation.\footnote{Here we only illustrate one mechanism by which non-repeating discontinuities can arise. We do not claim that this is the only such mechanism.} 
After computing discontinuities about $m_1^2=0$ and then $(m_1^2-m_2^2)=0$ by evaluating residues about the simple poles at the origin with respect to $z_1$ and $z_2$, the remaining integral is given by
\begin{equation}
        \text{Disc}_{(m_1^2-m_2^2)=0}\left(\text{Disc}_{m_1^2=0}\left( \mathcal{I}_{\text{hook}} \right)\right) \sim \int_{\nu_{12}} \frac{dz_3}{z_3} \left( z_3(m_1^2-m_2^2) - m_2^2(p^2-m_1^2+m_2^2) \right)^{\frac{D}{2}-2},
\end{equation}
where the vanishing cycle $\nu_{12}$ is shown in Figure~\ref{subfig:middle}. In this case, we see this cycle can be deformed into a residue contour about the remaining simple pole at the origin, as shown in Figure~\ref{subfig:right}. Thus, after computing a discontinuity with respect to $(m_1^2-m_2^2) = 0$, it is as if we have placed the third propagator on shell, even though reaching the singularity at $m_1 \to m_2$ does not require it. 

More formally, we can see why a second discontinuity with respect to $(m_1^2-m_2^2) = 0$ vanishes by constructing the vanishing cell, shown as $\tilde{\nu}_{12}$ in Figures~\ref{subfig:middle} and~\ref{subfig:right}. Namely, if we compute another discontinuity about $(m_1^2-m_2^2)=0$, the intersection number $\langle \tilde{\nu}_{12} | \nu_{12} \rangle =0$, which causes the right side of the Picard--Lefschetz formula to vanish. More generally, non-repeating discontinuities can be thought of as arising when the vanishing cycle and vanishing cell that correspond to a given discontinuity have a vanishing intersection number.

Before ending this section, let us quickly prove that non-repeating discontinuities cannot be of square root type. We will prove this by contradiction. Consider a Feynman integral $\mathcal{I}$ that has a singularity locus $\lambda=0$ of non-repeating type. First, we show that $\sqrt{\lambda}$ cannot appear in the prefactor of any polylogarithm, once the integral $\mathcal{I}$ has been evaluated in terms of iterated integrals. If this square root (or more generally, such a square root branch point) were present, we would be able to compute a maximal number of logarithmic discontinuities, to isolate just this algebraic prefactor. But the repeated discontinuity of (a rational function of) $\sqrt{\lambda}$ with respect to $\lambda=0$ is nonzero. Thus, $\sqrt{\lambda}$ cannot appear in these polylogarithmic prefactors.

Now let us assume the polylogarithmic part of $\mathcal{I}$ itself depends on $\sqrt{\lambda}$. We can then decompose the integral into its Galois-even (${\mathcal I}_+$) and Galois-odd (${\mathcal I}_-$) parts:
\begin{equation}
     {\cal I} = {\cal I}_+\left(\sqrt{\lambda}\right) + {\cal I}_-\left(\sqrt{\lambda}\right).
\end{equation}
The Galois-even part remains invariant under the replacement $\sqrt{\lambda}\to-\sqrt{\lambda}$, while the Galois-odd part changes sign. Thus, computing a discontinuity with respect to $\lambda$ yields
\begin{align}
    {\rm Disc}_\lambda {\cal I} &= \left[ {\cal I}_+\left(\sqrt{\lambda}\right) - {\cal I}_-\left(\sqrt{\lambda}\right) \right] - \left[{\cal I}_+\left(\sqrt{\lambda}\right) + {\cal I}_-\left(\sqrt{\lambda}\right)\right] \\
    &= - 2 {\cal I}_-\left(\sqrt{\lambda}\right).
\end{align}
However, by these same Galois properties, if we now compute a second discontinuity with respect to $\lambda=0$, the resulting expression will be non-zero:
\begin{equation}
    {\rm Disc}_\lambda {\rm Disc}_\lambda {\cal I} = 4 {\cal I}_-\left(\sqrt{\lambda}\right).
\end{equation}
Since this contradicts our original assumption, we conclude that non-repeating discontinuities cannot be of square root type.

\section{The Lefschetz Uniqueness Principle}
Now let us consider a slightly different situation, in which one finds that the number of point solutions to the critical point equations~\eqref{eq:critical-point-equations}
drops just by one in the $\lambda \to 0$ limit. This implies that there is a unique Lefschetz thimble that can play the role of a vanishing cycle in the Picard--Lefschetz formula. It immediately follows that a discontinuity with respect to $\lambda=0$ yields the same result, up to an integer proportionality factor, no matter which other discontinuities (or analytic continuations) we have computed first. In other words,
\begin{align}
    \text{Disc}_{\lambda=0}\cdots \mathcal{I} \ \, \propto \ \, \text{Disc}_{\lambda=0}\cdots \mathcal{I}\, ,
\end{align}
where the dots on either side of this equation represent different, arbitrary sequences of discontinuities and analytic continuations.

Importantly, as the above statement holds in any spacetime dimension, we can apply it to the $\epsilon$ expansion of our integral, where $\epsilon= \frac{4 - D}{2}$. If we do so, we end up relating different orders in the epsilon expansion, which suggests re-writing the above relation as\footnote{We thank Stefan Weinzierl for this observation.}
\begin{align}
    \frac{\text{Disc}_{\lambda=0}}{\epsilon} \frac{\text{Disc}_{\lambda_p=0}}{\epsilon}\cdots \frac{\text{Disc}_{\lambda_1=0}}{\epsilon} \mathcal{I} \ \, \propto \ \, \frac{\text{Disc}_{\lambda=0}}{\epsilon} \frac{\text{Disc}_{\lambda_q^\prime=0}}{\epsilon}\cdots \frac{\text{Disc}_{\lambda_1^\prime=0}}{\epsilon} \mathcal{I}\, ,
\end{align}
so that the transcendental weight of the two sides match order-by-order in the $\epsilon$ expansion, even when $p\neq q$ (where we have assumed all discontinuities are logarithmic, for simplicity). In practice, if $\lambda$ appears as a symbol letter in $\mathcal{S}(\mathcal{I})$, this means that the sequences of letters (or, more generally, the linear combinations of symbol words) that follow $\lambda$ must always be the same (including if these sequences of letters involve $\lambda$ itself). For example, if the only symbol term that involves $\lambda$ in the $\mathcal{O}(\epsilon^0)$ contribution to a two-loop integral $\mathcal{I}$ is
\begin{equation}
 a \otimes b\otimes \lambda \otimes d
\end{equation}
then we can deduce that $d$ is the only weight-one word that will ever follow $\lambda$ when the latter appears in the penultimate entry of $\mathcal{S}(\mathcal{I})$, at any order in $\epsilon$.

\section{Massless Steinmann Violation}
Finally, let us consider the Steinmann relations, which were originally proven for scattering amplitudes involving massive particles, but which have empirically been found to hold for massless amplitudes and integrals in most instances~\cite{Caron-Huot:2016owq}. It was recently shown that at one and two loops, all known violations can be traced back to solutions to the Landau equations that can be associated with one of the following minimal cuts (with minimal cuts as defined in~\cite{Hannesdottir:2025bss}):
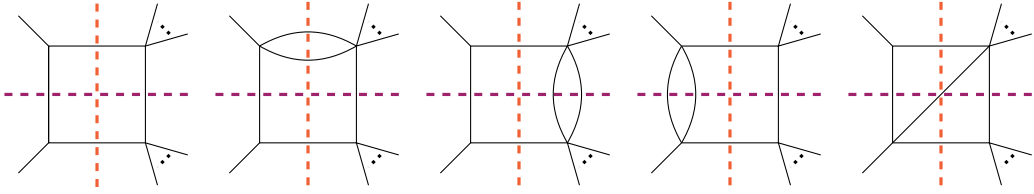
\begin{equation*}
\begin{tikzpicture}[scale=0.8]
        \coordinate (a) at (-.8,.8);
        \coordinate (b) at (.8,.8);
        \coordinate (c) at (.8,-.8);
        \coordinate (d) at (-.8,-.8);
        \coordinate (h) at (.8,0);

        \coordinate (a1) at (-1.3, 1.3);
        \coordinate (d1) at (-1.3, -1.3);
        \coordinate (b1) at (1, 1.5);
        \coordinate (b2) at (1.5, 1);
        \coordinate (c1) at (1, -1.5);
        \coordinate (c2) at (1.5, -1);

        \coordinate (cutL) at (-1.53, 0);
        \coordinate (cutR) at (1.53, 0);
        \coordinate (cutB) at (0,-1.53);
        \coordinate (cutT) at (0,1.53);
        
        \draw[] (a1) to (a);
        \draw[] (d1) to (d);
        \draw[] (b1) to (b);
        \draw[] (b2) to (b);
        \draw[] (c1) to (c);
        \draw[] (c2) to (c);

        \draw[dashed,thick,RedViolet,line width=1.2pt] (cutL) to (cutR);
        \draw[dashed,thick,RedOrange,line width=1.2pt] (cutB) to (cutT);

       \draw [dotted, line width=1.2pt] (1.2, 1) -- (1,1.2);
        \draw [dotted, line width=1.2pt] (1.2, -1) -- (1,-1.2);

        \draw[] (d) to node[midway, left] {}(a)
        to node[midway, above] {}(b)
        to node[midway, right] {}(c) 
        to node[midway, below] {}(d)
        to node[midway, right] {}(a);      
\end{tikzpicture}
\hspace{0.3cm}
\begin{tikzpicture}[scale=0.8] 
        \coordinate (a) at (-.8,.8);
        \coordinate (b) at (.8,.8);
        \coordinate (c) at (.8,-.8);
        \coordinate (d) at (-.8,-.8);
        \coordinate (a1) at (-1.3, 1.3);
        \coordinate (d1) at (-1.3, -1.3);
        \coordinate (b1) at (1, 1.5);
        \coordinate (b2) at (1.5, 1);
        \coordinate (c1) at (1, -1.5);
        \coordinate (c2) at (1.5, -1);
        
        \draw[] (a) to [out=-30,in=-150] (b);
        \draw[] (a) to [out=30,in=150](b);
        \draw[] (b) to (c) to (d) to (a);
        
        \draw[] (b1) to (b) to (b2);
        \draw[] (c1) to (c) to (c2);
        \draw[] (a1) to (a);
        \draw[] (d1) to (d);
        \draw [dotted, line width=1.2pt] (1.2, 1) -- (1,1.2);
        \draw [dotted, line width=1.2pt] (1.2, -1) -- (1,-1.2);

        \draw[RedOrange,thick,dashed,line width=1.2pt] (0,1.53) to (0,-1.53);
        \draw[RedViolet,thick,dashed,line width=1.2pt] (-1.53,0) to (1.53,0);
          
    \end{tikzpicture}\hspace{0.3cm}
    \begin{tikzpicture}[scale=0.8]
        \coordinate (a) at (-.8,.8);
        \coordinate (b) at (.8,.8);
        \coordinate (c) at (.8,-.8);
        \coordinate (d) at (-.8,-.8);
        \coordinate (a1) at (-1.3, 1.3);
        \coordinate (d1) at (-1.3, -1.3);
        \coordinate (b1) at (1, 1.5);
        \coordinate (b2) at (1.5, 1);
        \coordinate (c1) at (1, -1.5);
        \coordinate (c2) at (1.5, -1);
        
        \draw[] (c) to (d) to (a) to (b);
        \draw[] (b) to [out=-120,in=120] (c);
        \draw[] (b) to [out=-60,in=60] (c);
        
        \draw[] (b1) to (b) to (b2);
        \draw[] (c1) to (c) to (c2);
        \draw[] (a1) to (a);
        \draw[] (d1) to (d);
        \draw [dotted, line width=1.2pt] (1.2, 1) -- (1,1.2);
        \draw [dotted, line width=1.2pt] (1.2, -1) -- (1,-1.2);

        \draw[RedOrange,thick,dashed,line width=1.2pt] (0,1.53) to (0,-1.53);
        \draw[RedViolet,thick,dashed,line width=1.2pt] (-1.53,0) to (1.53,0);
          
    \end{tikzpicture}\hspace{0.3cm}
\begin{tikzpicture}[scale=0.8]
        \coordinate (a) at (-.8,.8);
        \coordinate (b) at (.8,.8);
        \coordinate (c) at (.8,-.8);
        \coordinate (d) at (-.8,-.8);
        \coordinate (a1) at (-1.3, 1.3);
        \coordinate (d1) at (-1.3, -1.3);
        \coordinate (b1) at (1, 1.5);
        \coordinate (b2) at (1.5, 1);
        \coordinate (c1) at (1, -1.5);
        \coordinate (c2) at (1.5, -1);
        
        \draw[] (a) to (b) to (c) to (d);
        \draw[] (d) to [out=120, in = -120] (a);
        \draw[] (d) to [out=60, in = -60] (a);
        
        \draw[] (b1) to (b) to (b2);
        \draw[] (c1) to (c) to (c2);
        \draw[] (a1) to (a);
        \draw[] (d1) to (d);
        \draw [dotted, line width=1.2pt] (1.2, 1) -- (1,1.2);
        \draw [dotted, line width=1.2pt] (1.2, -1) -- (1,-1.2);

        \draw[RedOrange,thick,dashed,line width=1.2pt] (0,1.53) to (0,-1.53);
        \draw[RedViolet,thick,dashed,line width=1.2pt] (-1.53,0) to (1.53,0);
          
    \end{tikzpicture}\hspace{0.3cm}
    \begin{tikzpicture}[scale=0.8]
        \coordinate (a) at (-.8,.8);
        \coordinate (b) at (.8,.8);
        \coordinate (c) at (.8,-.8);
        \coordinate (d) at (-.8,-.8);
        \coordinate (a1) at (-1.3, 1.3);
        \coordinate (d1) at (-1.3, -1.3);
        \coordinate (b1) at (1, 1.5);
        \coordinate (b2) at (1.5, 1);
        \coordinate (c1) at (1, -1.5);
        \coordinate (c2) at (1.5, -1);
        
        \draw[] (a) to (b) to (c) to (d) to (a);
        \draw[] (d) to (b);
        
        \draw[] (b1) to (b) to (b2);
        \draw[] (c1) to (c) to (c2);
        \draw[] (a1) to (a);
        \draw[] (d1) to (d);
        \draw [dotted, line width=1.2pt] (1.2, 1) -- (1,1.2);
        \draw [dotted, line width=1.2pt] (1.2, -1) -- (1,-1.2);

        \draw[RedOrange,thick,dashed,line width=1.2pt] (0,1.53) to (0,-1.53);
        \draw[RedViolet,thick,dashed,line width=1.2pt] (-1.53,0) to (1.53,0);
          
    \end{tikzpicture}
\end{equation*}
Similar solutions to the Landau equations that can give rise to Steinmann violation can be constructed in conjunction with higher-loop cuts of the form:
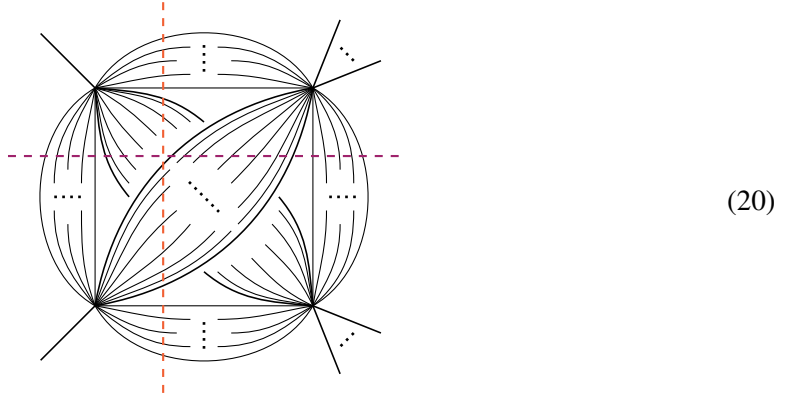
\begin{equation}
\begin{gathered} \label{eq:connected_minimal_cuts}
        \begin{tikzpicture}[scale=1.8]
        \coordinate (a) at (-.8,.8);
        \coordinate (b) at (.8,.8);
        \coordinate (c) at (.8,-.8);
        \coordinate (d) at (-.8,-.8);
        \coordinate (h) at (.8,0);

        \coordinate (a1) at (-1.2, 1.2);
        \coordinate (d1) at (-1.2, -1.2);
        \coordinate (b1) at (1.2, 1.2);
        \coordinate (c1) at (1.2, -1.2);

        \coordinate (i2) at (-0.3, 1.3);
        \coordinate (j2) at (0.3, 1.3);

        \draw[] (b) to (c);
        \draw[] (d) to (a);
        \coordinate (e1) at (1.2, 0.2);
        \coordinate (f1) at (1.2, -0.2);
        \coordinate (e2) at (1.3, 0.3);
        \coordinate (f2) at (1.3, -0.3);
        
        \draw[] (b) to [out=-80, in=90] (0.9,0.1);
        \draw[] (0.9,-0.1) to [out=-90, in=80] (c);
        \draw[] (b) to [out=-60, in=90] (1,0.2);
        \draw[] (1,-0.2) to [out=-90, in=60] (c);
        \draw[] (b) to [out=-40, in = 90] (1.1,0.1);
        \draw[] (1.1,-0.1) to [out=-90, in=40] (c);
        \draw[] (b) to [out=-30, in = 30] (c);
        
        \coordinate (g1) at (-1.2, 0.2);
        \coordinate (h1) at (-1.2, -0.2);
        \coordinate (g2) at (-1.3, 0.3);
        \coordinate (h2) at (-1.3, -0.3);

        \draw[] (a) to [out=-100, in = 90] (-0.9,0.1);
        \draw[] (-0.9,-0.1) to [out=-90, in = 110] (d);
        \draw[] (a) to [out=-120, in = 90] (-1,0.2);
        \draw[] (-1,-0.2) to [out=-90, in = 120] (d);
        \draw[] (a) to [out=-130, in = 90] (-1.1,0.1);
        \draw[] (-1.1,-0.1) to [out=-90, in=130] (d);
        \draw[] (a) to [out=-150, in=150] (d);

        \draw [dotted, line width=1pt] (0.92, 0) -- (1.12,0);
        \draw [dotted, line width=1pt] (-1.1, 0) -- (-0.9,0);
        
        \draw[] (a) to (b);
        \draw[] (a) to [out=10, in = 180] (-0.1, 0.9);
        \draw[] (0.1, 0.9) to [out=0, in = 170] (b);
        \draw[] (a) to [out=30, in= 180] (-0.2, 1);
        \draw[] (0.2, 1) to [out=0, in=150] (b);
        \draw[] (a) to [out=40, in=180] (-0.1, 1.1);
        \draw[] (0.1, 1.1) to [out=0, in=140] (b);
        \draw[] (a) to [out=60, in=120] (b);

        \draw [dotted, line width=1pt] (0, 0.92) -- (0,1.12);

        \draw[] (d) to (c);
        \draw[] (d) to [out=-10, in=180] (-0.1, -0.9);
        \draw[] (0.1, -0.9) to [out=0, in=-170] (c);
        \draw[] (d) to [out=-30, in=180] (-0.2, -1);
        \draw[] (0.2, -1) to [out=0, in=-150] (c);
        \draw[] (d) to [out=-40, in=180] (-0.1,-1.1);
        \draw[] (0.1,-1.1) to [out=0, in=-140] (c);
        \draw[] (d) to [out=-60, in=-120] (c);

        \draw[dotted, line width=1pt] (0,-0.92) -- (0,-1.12);

        \draw [dotted, line width=1pt] (-0.11, 0.11) -- (0.11,-0.11);
        \draw[line width=0.6pt] (d) to [out=80, in=-170] (b);
        \draw[line width=0.6pt] (d) to [out=10, in=-100] (b);
        \draw[] (d) to [out=50, in=-140] (-0.2, -0.1);
        \draw[] (0.1,0.2) to [out=45, in=-140] (b);
        \draw[] (d) to [out=40, in=-130] (-0.1, -0.2);
        \draw[] (0.2,0.1) to [out=45, in=-130] (b);
        \draw[] (d) to [out=60, in=-137] (-0.25, 0.08);
        \draw[] (-0.08, 0.25) to [out=45, in=-150] (b);
        \draw[] (d) to [out=30, in=-133] (0.08, -0.25);
        \draw[] (0.25, -0.08) to [out=45, in=-120] (b);
        \draw[] (d) to [out=70, in=-136] (-0.25,0.2);
        \draw[] (-0.2,0.25) to [out=44, in=-160] (b);
        \draw[] (d) to [out=20, in=-134] (0.2,-0.25);
        \draw[] (0.25,-0.2) to [out=46, in=-110] (b);

         \draw[line width=0.6pt] (a) to [out=-5, in=140] (0,0.55);
         \draw[line width=0.6pt] (a) to [out=-85, in=130] (-0.55,0);
         \draw[line width=0.6pt] (0.55,0) to [out=-50, in=95] (c);
         \draw[line width=0.6pt] (0,-0.55) to [out=-40, in=175] (c);
         \draw[] (a) to [out=-10, in=138] (-0.1, 0.55);
         \draw[] (a) to [out=-80, in=132] (-0.55, 0.1);
         \draw[] (0.1, -0.55) to [out=-42, in=170] (c);
         \draw[] (0.55, -0.1) to [out=-48, in=100] (c);
         \draw[] (a) to [out=-15, in=137] (-0.2, 0.5);
         \draw[] (a) to [out=-75, in=133] (-0.5,0.2);
         \draw[] (0.2,-0.5) to [out=-43, in=165] (c);
         \draw[] (0.5,-0.2) to [out=-47, in=105] (c);
         \draw[] (a) to [out=-30, in=136] (-0.35, 0.45);
         \draw[] (a) to [out=-60, in=134] (-0.45, 0.35);
         \draw[] (0.35, -0.45) to [out=-46, in=155] (c);
         \draw[] (0.45, -0.35) to [out=-44, in=115] (c);
         
        \coordinate (a1) at (-1.2, 1.2);
        \coordinate (d1) at (-1.2, -1.2);
        \coordinate (b1) at (1.2, 1.2);
        \coordinate (c1) at (1.2, -1.2);
        
        \draw[line width=0.6pt] (1,1.3) to (b);
        \draw[line width=0.6pt] (1.3,1) to (b);
        \draw[line width=0.6pt] (1,-1.3) to (c);
        \draw[line width=0.6pt] (1.3,-1) to (c);

        \draw[line width=0.6pt] (a1) to (a);
        \draw[line width=0.6pt] (d1) to (d);
        
        \draw[dotted, line width=1pt] (1.1,1) to (1,1.1);
        \draw[dotted, line width=1pt] (1.1,-1) to (1, -1.1);

        \draw[RedViolet,thick,dashed] (-1.44,0.3) to (1.44,0.3);
        \draw[RedOrange,thick,dashed] (-.3,-1.44) to (-.3,1.44);
\end{tikzpicture}
\end{gathered}
\end{equation}
This makes it possible to formulate a graphical rule for when we expect the Steinmann relations to be violated~\cite{Hannesdottir:2025bss}.\footnote{Note that this statement just applies to the first two discontinuities of an integral, not any adjacent pair of discontinuities, as considered for instance in~\cite{Drummond:2017ssj,Caron-Huot:2019bsq}.} Namely, if the minimal cut for a given pair of momentum channels takes the form~\eqref{eq:connected_minimal_cuts}, then it is expected that Steinmann will be violated (at some order in $\epsilon$). This provides us with additional data on what properties should be built into the ansatz for massless Feynman integrals. 

\section{Conclusions}

We have presented several new constraints on the sequential discontinuities of Feynman integrals that can be derived using ideas from Picard--Lefschetz theory. In particular, the Lefschetz uniqueness principle and the identification of non-repeating discontinuities constitute new types of all-orders constraints that can be efficiently and algorithmically deduced for individual Feynman integrals~\cite{to_appear}. Together, these results enlarge the set of available tools in the Landau bootstrap and enrich our conceptual understanding of Feynman integrals. At the same time, we expect that still more powerful constraints remain to be uncovered using these homological methods.

\newpage
\bibliographystyle{apsrev4-1}
\bibliography{refs}

\end{document}

%% file: tikz_figs/1a.tex
\begin{tikzpicture}
    \draw[line width=0.85pt] (-1, 0.) -- (0.5, 0.866025);
    \draw[line width=2pt] (-1, 0.) -- (0.5, -0.866025);
    \draw[line width=2pt] (0.5, -0.866025) -- (0.5, 0.866025);
    \draw[line width=0.85pt] (-1.49666, -0.1) -- (-1., 0.) -- (-1.49666, 0.1);
    \draw[line width=0.85pt] (0.5, -0.866025) -- (0.75, -1.29904);
    \draw[line width=0.85pt] (0.5, 0.866025) -- (0.75, 1.29904);
    
    \node at (-1.85,0.0) {\small\textit{$p$}};
    \node at (-0.4,0.7) {\small\textit{$\alpha_3$}};
    \node at (-0.4,-0.75) {\small\textit{$\alpha_2$}};
    \node at (-0.2,-0.2) {\small\textit{$m_2$}};
    \node at (0.9,0.0) {\small\textit{$\alpha_1$}};
    \node at (0.22,0.2) {\small\textit{$m_1$}};
\end{tikzpicture}

%% file: tikz_figs/1b.tex
\begin{tikzpicture}[scale=1, >=Stealth]

    \draw[->, thick] (-1.5, 0) -- (2.7, 0) ;
    \draw[->, thick] (0, -1.5) -- (0, 1.5) ;

    \draw[thick] (2.05,1.20) -- (2.50,1.20);
    \draw[thick] (2.05,1.20) -- (2.05,1.60);
    \node at (2.30,1.40) {$z_3$};

    \filldraw[black] (0,0) circle (2.0pt);
    
    \draw[lightgray,thick] (0.9,-0.1) -- (1.1,0.1);
    \draw[lightgray,thick] (0.9,0.1) -- (1.1,-0.1);
    
    \draw[
        lightgray,
        thick,  
        decoration={zigzag, amplitude=1.5pt, segment length=4pt, pre=moveto, pre length=2pt}, 
        decorate
    ] (1,0) -- (2.55,0);

    \node at (1.9,0.4) {\footnotesize\textit{$\nu_{12}$}};
    \draw[very thick, teal] (1.0,0.2) -- (2.6,0.2);
    \draw[very thick, teal] (1.0,0.2) .. controls (0.75,0.2) and (0.75,-0.2) .. (1.0,-0.2);
    \draw[very thick, teal] (1.0,-0.2) -- (2.6,-0.2);
    \draw [very thick, teal,->,-{To[scale=0.7]}] (1.8,0.2) -- (1.5,0.2);
    \draw [very thick, teal,->,-{To[scale=0.7]}] (1.5,-0.2) -- (1.8,-0.2);

    \draw[very thick, purple] (1.0,0.0) -- (2.55,0.0);
    \draw[very thick, purple,->,-{To[scale=0.7]}] (1.5,0.0) -- (1.9,0.0);

\end{tikzpicture}

%% file: tikz_figs/1c.tex
\begin{tikzpicture}[scale=1, >=Stealth]

    \draw[->, thick] (-1.5, 0) -- (2.7, 0) ;
    \draw[->, thick] (0, -1.5) -- (0, 1.5) ;

    \draw[thick] (2.05,1.20) -- (2.50,1.20);
    \draw[thick] (2.05,1.20) -- (2.05,1.60);
    \node at (2.30,1.40) {$z_3$};

    \filldraw[] (0,0) circle (2.0pt);

    \node at (-0.4,0.3) {\footnotesize\textit{$\nu_{12}$}};
    \draw[
        very thick,
        teal,
        postaction={
            decorate,
            decoration={
                markings,
                mark=at position 0.5 with {\arrow{Computer Modern Rightarrow}}
            }
        }
    ] (0,0) circle [radius=0.2];

    \draw[lightgray,thick] (0.9,-0.1) -- (1.1,0.1);
    \draw[lightgray,thick] (0.9,0.1) -- (1.1,-0.1);

    \draw[
        lightgray,
        thick,  
        decoration={zigzag, amplitude=1.5pt, segment length=4pt, pre=moveto, pre length=2pt}, 
        decorate
    ] (1,0) -- (2.55,0);

    \node at (1.9,-0.3) {\footnotesize\textit{$\tilde{\nu}_{12}$}};
    \draw[very thick, purple] (1.0,0.0) -- (2.55,0.0);
    \draw[very thick, purple,->,-{To[scale=0.7]}] (1.5,0.0) -- (1.9,0.0);

\end{tikzpicture}